\documentclass[a4paper,11pt]{article}
\pdfoutput=1 

\usepackage{jheppub} 

\usepackage{graphicx} 
\usepackage{epstopdf} 
\usepackage{slashed}
\newcommand{\GeV}{\,\mathrm{GeV}}

\newcommand{\mttwo}{M_{T2}}

\begin{document}

\title{A faster, high precision algorithm for calculating symmetric and asymmetric $\mttwo$}

\author{Colin H. Lally}
\affiliation{Cranbrook School, Waterloo Road, Kent, TN17 3JD, United Kingdom}


\emailAdd{lallyc@cranbrook.kent.sch.uk}

\abstract{
A new algorithm for calculating the stransverse mass, $\mttwo$, in either symmetric or asymmetric situations has been developed which exhibits good stability, high precision and quadratic convergence for the majority of the $\mttwo$ parameter space, leading to up to a factor of ten increase in speed  compared to other $\mttwo$ calculators of comparable precision. This document describes and validates the methodology used by the algorithm, and provides comparisons both in terms of accuracy and speed with other existing implementations.}

\maketitle

\section{Introduction}

\label{sec:lalintro}

The avid reader of papers on $\mttwo$ phenomenology might be aware that a paper with a similar-sounding title and abstract was published a few months ago by Lester and Nachman\cite{Lester:2014yga}. Their motivation (to produce a high precision calculator that does not suffer from inaccuracies in certain difficult regions of the $\mttwo$ parameter space) is a similar one to this paper, and the underlying methodologies used share a remarkably similar starting point (as both rely on the same key principle\footnote{It transpires that this author and one of the authors of \cite{Lester:2014yga} (CGL) had independently come across similar work involving the matrix addition of conics, the key idea behind these two latest implementations.}). Where the two papers differ, however, is that the key motivation behind the work presented here was to produce as {\em fast} an algorithm for evaluating $\mttwo$ as possible (whilst still of course producing results to a similar high level of precision); this has led to quite a different methodology when compared with \cite{Lester:2014yga} that calculates $\mttwo$ at least as accurately (if not more accurately - see section \ref{ssec:algaccuracy}), and which uses arguably a more straightforward implementation in terms of the coding involved\footnote{However possibly not as {\em robust} an implementation as \cite{Lester:2014yga} - see section \ref{ssec:fastermethod}.}. 

\section{$\mttwo$ and its computation}
\label{sec:calculatingMT2}

The stransverse mass, $\mttwo$, \cite{Lester:1999tx} is a now well-established kinematic technique for hadron collider events that, in its simplest guise, uses the missing transverse momenta of what is assumed to be two invisible (for example hypothetical R-parity conserving supersymmetric) daughter particles, along with the transverse momenta of the associated visible particles/jets, to establish a maximal lower-bound on the mass of the assumed pair-produced parent particles that gave rise to the visible and invisible particles. In its original incarnation, the masses of the two invisible daughter particles were assumed to be identical (the ``symmetric'' case), however $\mttwo$ can be trivially extended to (and is being increasingly used in) the ``asymmetric'' case, where the two daughter masses are assumed to be different. Properties and generalisations of $\mttwo$ have been investigated extensively \cite{Allanach:2000kt,Barr:2002ex,Barr:2003rg,Lester:2007fq,Cho:2007qv,Gripaios:2007is,Barr:2007hy,Cho:2007dh,Ross:2007rm,Nojiri:2007pq,Tovey:2008ui,Cho:2008cu,Serna:2008zk,Barr:2008ba,Cho:2008tj,Burns:2008va,Barr:2008hv,Barr:2009mx,Barr:2009jv,Polesello:2009rn,Kim:2009si,Konar:2009wn,Konar:2009qr,Baringer:2011nh,Walker:2013uxa,Cho:2014naa}
and the variable has been used in many experimental searches, especially for supersymmetric particles (see for example \cite{daCosta:2011qk,Collaboration:2012ida,c:2012gg,cm:2012jx,Weber:2012fa,Aad:2014kra}).

In the general case $\mttwo$ must be solved numerically\footnote{In some special cases analytic expressions do exist; for example in the ``fully massless case'', where the visible and invisible daughter particles are all massless, it has been shown \cite{Lester:2011nj} that $\mttwo$ can be calculated from the roots of a quartic equation (see also \cite{Lally:2012uj} for further discussions of, and a fast, fully-analytic implementation for, $\mttwo$ in the fully massless case.)}, and although various algorithms exist for doing so, the important ``ellipse bisection'' methodology and associated algorithm developed by Cheng and Han in 2008 \cite{Cheng:2008mg} (who, following the notation of \cite{Lester:2014yga}, will be referred to here as ``CH'') is regarded as the simplest and most efficient way to calculate $\mttwo$ (note a useful library of implementations, including the CH algorithm, can be found at \cite{oxbridgeStransverseMassLibrary}). In the CH implementation, kinematic constraints bound the ``trial'' mass values for $\mttwo$ into two (one for each ``side'' of an event) conic regions  (elliptical when the visible daughter particles are massive, parabolic when they are massless) in the invisible particles' phase space, and computation of $\mttwo$ is reduced to finding the smallest trial mass that ensures the two conic regions intersect i.e. the smallest trial mass that could physically have given rise to both sides of the event. The method for finding the intersection involves bisecting an interval, a robust computational method that guarantees an answer to a requested level of precision with linear convergence. However, as pointed out in \cite{Lester:2014yga}, the CH implementation suffers from accuracy problems when trying to calculate $\mttwo$ in the areas of its parameter space where events have very light, but nonetheless massive, daughter particles as opposed to massless ones; that is cusp regions of parameter space where the conic boundaries crossover from elliptical to parabolic (a parabola being an ellipse in the limit of infinitely large axes). It should be pointed out that this is not an error in the CH algorithm as such; any implementation for calculating $\mttwo$ has to find a way to deal with these problematic cusp regions, and CH decided the simplest way was to have an arbitrary minimum mass below which the event was regarded as massless, thus leading to reduced accuracy for this type of event. A detailed discussion of the various problematic cusp regions can be found in Walker's excellent tour de force on $\mttwo$ computation \cite{Walker:2013uxa}; Walker's paper also provides an all-encompassing algorithm for calculating $\mttwo$ reliably in a variety of ``difficult'' situations, however as noted by \cite{Lester:2014yga}, due to the ensuing complexity of the algorithm it is considerably slower than the CH implementation.

The recent release of the algorithm developed by Lester and Nachman detailed in \cite{Lester:2014yga} (we will refer to this going forward as the ``LN'' implementation) is an important one. Firstly it is designed to be robust in all regions of $\mttwo$ parameter space; it accomplishes this feat using two techniques: the first is that it represents the CH kinematic constraint equation for each ``side'' of an event by a generalised conic equation (technically this is in matrix form); this has the advantage of allowing the underlying geometry of the event to move seamlessly from elliptical to parabolic (as the masses of the invisible daughter particles approach zero) in a smooth manner; the second, and this is the key achievement, is that the algorithm does not look for an intersection of the perimeters of the two bound conical regions (as per previous implementations), but instead looks for an intersection of the bound regions themselves (what LN refer to as the ``interiors'' of the conics). This ensures that there is only ever one solution, that is when the bound regions first touch.\footnote{Note this also allows so-called ``unbalanced'' situations, where one ellipse is entirely bounded by the other (and which in existing implementations requires a separate computational check), to be found automatically as in this case the ``interiors'' of the conics \texttt{always} touch (leading for example to a value of $\mttwo$ of exactly zero in the fully massless case).} Secondly, the LN implementation can compute both symmetric {\it and} asymmetric events whereas, for purely historical reasons, the CH implementation can only be used in symmetric situations.\footnote{Indeed some less than successful attempts to modify CH's algorithm to deal with asymmetric events (see Walker's Table 1 in \cite{Walker:2013uxa}) was another key motivating factor behind the LN implementation.} The LN implementation is still a bisection method (and thus still linear in convergence) and so it should allow $\mttwo$ to be calculated for every physical event to a high precision. Even better, the LN implementation is about two times faster than the CH method to the typical precision of the CH algorithm ($\sim0.002 \GeV$) and only at worst about 50\% slower when calculating $\mttwo$ to a precision of $\sim10^{-12}\GeV$.

\section{Route to a faster algorithm}
\label{sec:fasteralgorithm}
\subsection{What is meant by a ``fast'' algorithm?}
\label{ssec:defoffast}

As the CH and LN implementations use a bisection method, both algorithms converge linearly and thus the computational time is proportional to the precision required (LN refer to a proportional relationship between computational time, $\tau$, and precision, $D$, in an algorithm i.e. $\tau \propto D$, as a ``fast'' method).\footnote{Of course the actual computational time depends as much on making as accurate a guess as one can of upper and lower limits for $\mttwo$ in a specific event (and thus giving as small a range to be bisected as possible) as it does on the linear convergence of the method. For example a clever addition to the LN algorithm, the so-called ``deci-section optimisation'' feature, gives, according to its authors, roughly a factor of three increase in speed for $\mttwo$ values near an event's kinematic minimum.} What is proposed here is an algorithm which exhibits mainly quadratic, but at worst superlinear, convergence in finding $\mttwo$ to a similarly high precision as the LN implementation; this greater convergence rate gives a computational speed which is up to ten times faster than any known existing methods for evaluating $\mttwo$.

\subsection{The faster methodology}
\label{ssec:fastermethod}

As per the work of CH, evaluating $\mttwo$ is reduced to finding the intersection of two ellipses (or parabolae, or combination of the two). To do this we will use a method due to \cite{wang2001algebraic}; although originally developed for ellipsoid intersection\footnote{Interestingly this method was developed in the context of robotics and virtual reality environments where the intersection of ellipsoids is a computationally efficient way to detect the collisions of 3D entities (represented as a combination of ellipsoids) with each other and with their surroundings.}, this was then extended to ellipses culminating in the lemmas and key theorem (Theorem 6) derived in \cite{choi2006continuous}; although \textit{only} proven for ellipses, there is no reason to believe the theorem should not be applicable to parabolae intersection as well, and so going forward we will refer to the theorem as if it was derived for general conics (see Appendix \ref{sec:validityconics} for a detailed analysis of why the theorem should be applicable to parabolae as well as to ellipses).

We begin by defining the equation of a conic region $\mathcal{P}$ for side one of an event (note in the following we ignore {\it hyperbolic} conic regions as these have no relevance for the computation of $\mttwo$):

\begin{equation}
\label{eq:conicP}
\mathcal{P} : A_p x^2 + B_p xy + C_p y^2 + D_p x + E_p y + F \leq 0
\end{equation}

Where for convenience the transverse momentum coordinates for side one $(p_{1x}, p_{1y})$ are written simply as $(x ,y)$. In matrix form this can be rewritten as $X^T P X \leq 0$, where $X^T = (x, y, 1)$ and $P = [P_{i, j}]$ is the following $3 \times 3$ real symmetric matrix

\begin{equation}
P = 
\left(
\begin{matrix} 
 A_p&\frac{B_p}{2}&\frac{D_p}{2}\\\frac{B_p}{2}&C_p&\frac{E_p}{2}\\\frac{D_p}{2}&\frac{E_p}{2}&F_p 
\end{matrix}
\right)
\label{eq:matrixP}
\end{equation}

Note $\mathcal{P}$ would represent an {\it elliptical} region if $det(P_{2,2}) > 0$ ($P_{2,2}$ being the $2 \times 2$ submatrix of $P$ with the 3rd row and 3rd column removed), and it would represent a {\it parabolic} region if  $det(P_{2,2}) = 0$ (and would be a {\it hyperbolic} region if $det(P_{2,2}) < 0$). Similarly the conic region for side two of the event in matrix form is defined as $\mathcal{Q} : X^T Q X \leq 0$. 

Given these two conic regions, the cubic polynomial $f(\lambda) = det(\lambda P - Q)$ (where $\lambda$ is some arbitrary factor) is called the {\it characteristic polynomial} and

\begin{equation}
\label{eq:chareq}
f(\lambda) = det(\lambda P - Q) = 0
\end{equation}

the {\it characteristic equation} of $\mathcal{P}$ and $\mathcal{Q}$. The key result (Theorem 6) of \cite{choi2006continuous} upon which the new implementation presented here (and indeed the implementation of LN) is based is as follows: given two conic regions $\mathcal{P} : X^T P X \leq 0$ and $\mathcal{Q} : X^T Q X \leq 0$

\begin{enumerate}
\item $\mathcal{P}$ and $\mathcal{Q}$ touch externally if and only if $f(\lambda) = 0$ has a negative real double root;
\item $\mathcal{P}$ and $\mathcal{Q}$ are separate if and only if $f(\lambda) = 0$ has two distinct negative real roots.
\end{enumerate}

\begin{figure}
\centering
\includegraphics[width=15.48cm]{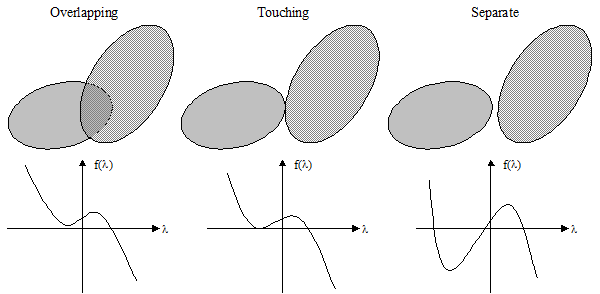}
\caption{Two elliptical regions and their characteristic polynomial $f(\lambda)$. Overlapping iff $f(\lambda) = 0$ has no negative real root, touching externally iff $f(\lambda) = 0$ has a double negative real root, and separate iff $f(\lambda) = 0$ has two distinct negative real roots. The single positive root is also evident in each case (reproduced from \cite{choi2006continuous}).}
\label{fig:EllipsesvsCharEq}
\end{figure}

It is also the case (see Lemma 1 of \cite{choi2006continuous}) that in both these cases (and the case where the conic regions overlap and there are no negative real roots - see  figure~\ref{fig:EllipsesvsCharEq}) the third root of $f(\lambda) = 0$ must be positive (and real).

These properties of the characteristic equation $f(\lambda) = 0$ lead to specific relationships between the coefficients of $f(\lambda)$ and it is these relationships which are utilised in the LN implementation to find the initial intersection point of two conic regions. The new implementation presented here however takes a different approach.

Before outlining the new approach, it is worth briefly reminding ourselves of some key features of CH's kinematically bound regions:

\begin{enumerate}
\item These regions correspond to kinematically allowed transverse momenta for the two invisible particles of an event, taking into account: an assumed parent particle mass, the energy and momenta of the associated visible particles/jets, the assumed masses of the invisible particles, and the event's overall missing transverse momentum. As the assumed parent mass is increased, the kinematically allowed transverse momenta regions increase. $\mttwo$ is the minimum parent particle mass that satisfies the kinematic constraints of {\it both} sides of the event, and hence will manifest as an initial intersection of the two allowed regions in the invisible transverse momentum coordinate system of one side of an event;
\item In an elliptical scenario, the allowed region begins as a point at a parent mass equal to the sum of the visible and assumed invisible mass (the kinematic minimum) and expands from this starting point as the assumed parent mass is increased; in a parabolic scenario, the allowed region begins as a straight ray (a line going from a fixed point going out to infinity in some direction) at the same kinematic minimum parent mass, and unfolds on either side of the ray into a parabola as the parent mass increases, with the vertex of the parabola at the original fixed point of the ray.
\end{enumerate}

The above is an embarrassingly brief description of these conic regions, the reader is reminded that an excellent and detailed description can be found in Walker's paper \cite{Walker:2013uxa}. The point of this brief description is to make clear that, as the conic regions are obviously functions of the trial parent mass, the resulting characteristic equation is also a function of the trial parent mass (as well as $\lambda$), this dependence being found in the coefficients of the characteristic equation. We will thus redefine the characteristic polynomial as $f(\lambda; \delta)$, where $\delta$ is defined as:

\begin{equation}
\label{eq:deltafunction}
\delta = \frac{\mu_Y^2 -\mu_{N}^2 - m_1^2}{2E_1^2}
\end{equation}

\begin{figure}
\centering
\includegraphics[width=9cm]{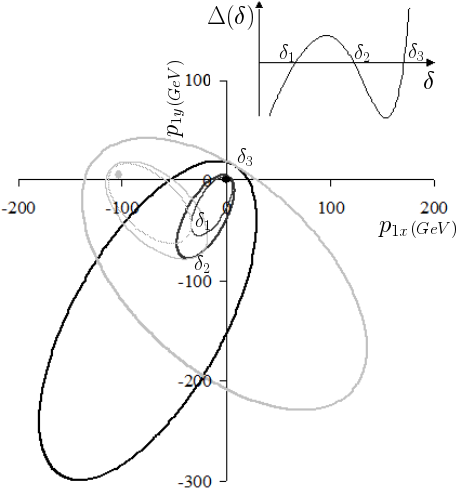}
\caption{Depicting a typical situation where the descriminant function, $\Delta(\delta)$, has three positive roots, $\delta_1$, $\delta_2$ and $\delta_3$. These correspond to $\mu_Y$ values where the two ellipses intersect: the two ellipses initially intersect at $\delta_1$ corresponding to the lowest positive root of the discriminant; for larger trial masses, $\mu_Y$, the ellipses expand and the two ellipses intersect on their boundaries at two further locations, $\delta_2$ and $\delta_3$.}
\label{fig:DiscriminantRoots}
\end{figure}

Here $\mu_Y$ is the ``trial'' parent mass, $\mu_{N}$ the invisible mass, and $m_1$ and $E_1$ the mass and energy of the most massive or, if massless, the most energetic side of the event.\footnote{These quantities are the same as those used in the CH implementation and are also used in the implementation proposed here. Note also that the quantity $\delta$ is equivalent to Walker's quantity $\Gamma$ in the limit of his one-step decay topology.}  As mentioned above, the key situation is where the two conic regions touch externally and this is where $f(\lambda; \delta) = 0$ has a multiple negative root. It is a well-established fact that the discriminant, $\Delta$, of a polynomial with multiple roots equals zero. The discriminant of the cubic characteristic polynomial is calculated from its coefficients, and thus the discriminant is a function of the trial parent mass. Finding $\mttwo$ then reduces to finding the lowest positive $\delta$ value (and hence from Equation \eqref{eq:deltafunction} the lowest $\mu_Y$ value) that ensures $\Delta = 0$ i.e. finding the lowest positive root of $\Delta = 0$. It is worth noting that the discriminant can and does have larger positive roots; in fact in the elliptical scenario the discriminant is an eight-order polynomial in $\delta$ (which reduces to a quartic polynomial in the parabolic, i.e. massless, scenario\footnote{Thus reconfirming the result first derived in \cite{Lester:2011nj} that $\mttwo$ in the fully massless case can be found analytically by solving a quartic equation}). Any larger positive roots correspond to trial masses where the boundary of one conic region touches the other's boundary beyond the initial intersection of the exteriors (see figure~\ref{fig:DiscriminantRoots}).

Reducing the computation of $\mttwo$ to a simple root-finding procedure (though one where it is always clear exactly \textit{which} is the correct root i.e. it is the delta value which ensures the characteristic equation has a double negative root) makes sense because for a smooth, differentiable function such as the discriminant, root-finding using numerical techniques is a well-established field with many efficient algorithms available. One of the oldest, simplest and fastest is the Newton-Raphson method (which has quadratic convergence, see for example \cite{ypma1995historical}), another is the Regula Falsi method (which has guaranteed superlinear convergence as long as a \textit{modified} algorithm is chosen\footnote{A number of modified methods are available and have been tested for this implementation; however given the straight-forward nature of the discriminant function, one of the simpler modifications, called the \textit{Pegasus} method, was found to be just as efficient as more complicated modifications.}, see for example \cite{ford1995improved}). Both are classic methods but both have some well-known limitations. An important feature of either is to ensure an iteration begins from as near to a root as possible, and fortunately there are some simple techniques for setting bounds on $\mttwo$ (see for example \cite{Walker:2013uxa} for a detailed discussion of upper and lower bounds on $\mttwo$). By finding sensible bounds on the trial parent mass prior to starting the iteration, $\mttwo$ can be computed with around ten iterations using the Newton-Raphson method for the vast majority of events to a precision of $\sim10^{-12}\GeV$. However in some cases (it is difficult to be precise but perhaps around 1\%) the Newton-Raphson iteration either finds one of the higher roots (identified due to the characteristic equation's lack of a double negative root) or simply doesn't converge particularly quickly; in this situation the algorithm needs to use a slower (bisection-like) method to set revised bounds for the trial parent mass before it can re-compute the correct root. As the Newton-Raphson method is not guaranteed to find a root between the two new bounds (one of its flaws unfortunately), this time the algorithm uses the Regula Falsi method (which is guaranteed to find the correct root albeit with only superlinear convergence). This potential need to switch between two iteration methods evidences the fact that this new implementation requires certain checks that the LN implementation does not; although it has been tested successfully on millions of different types of events it is still possible that there could be some unusual event that might cause the new algorithm to fail to find the correct $\mttwo$ value. These events would hopefully be extremely rare at worst, but, for this reason, the LN implementation should still be regarded as the {\it most} robust method available for calculating $\mttwo$.

\section{Algorithm validation}
\label{sec:algorithmvalidation}
\subsection{Accuracy}
\label{ssec:algaccuracy}

As discussed previously, the CH implementation, whilst a significant step-forward in $\mttwo$ computation, does suffer from accuracy issues as well as being limited to symmetric $\mttwo$ events, and so the LN implementation is to be regarded as currently not only the most robust but also the most \textit{efficient} algorithm available for the \textit{high precision} computation of $\mttwo$. Thus to validate the accuracy of this new proposed implementation, it was felt sufficient to only compare computed $\mttwo$ values to those obtained using the LN implementation.

\begin{figure}
\input{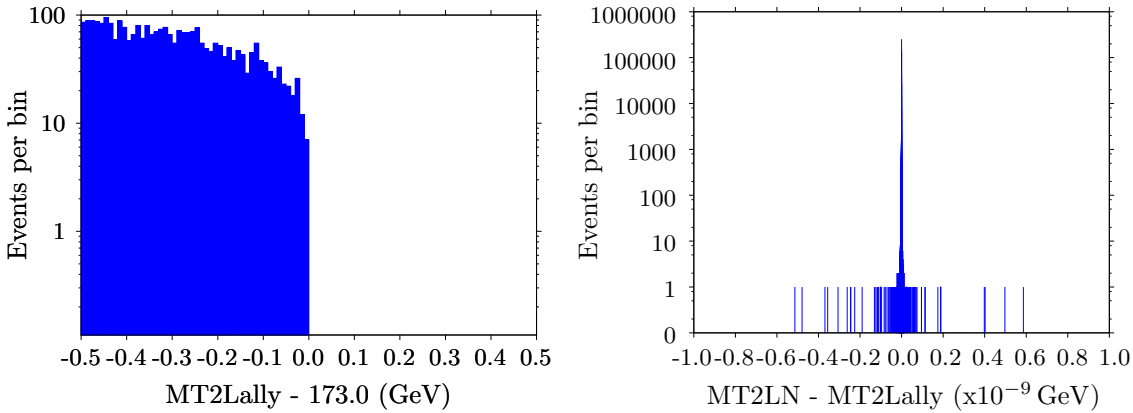}
\caption{The left hand plot shows the kinematic endpoint, $\pm0.5 \GeV$, of a ${t}\bar{t}$ $\mttwo$ distribution for one million events. The distribution would be expected to stop at exactly 0 (having subtracted the top mass) which it does. The right hand plot shows the difference between these $\mttwo$ values calculated with the proposed implementation and those due to LN, demonstrating how well the two algorithms agree - for the vast majority of events ($> 97\%$) the agreement is better than $\sim10^{-12}\GeV$.}
\label{fig:lesterlallyTopdiff}
\end{figure}

In the first scenario a toy MC that pair-produces on-shell particles was used to generate millions of truth-level ${t}\bar{t}$ events; $\mttwo$ was then reconstructed from the (visible) ${b}$-jet and lepton masses and momenta and the (invisible) neutrino missing momenta. An accurate implementation for computing $\mttwo$ would expect to produce values that have a sharp cut-off at the top mass. The left-hand plot in figure~\ref{fig:lesterlallyTopdiff} shows the upper kinematic end-point of the resulting distribution having subtracted the top mass (assumed to be exactly 173.0$\GeV$), demonstrating the expected sharp cut-off at zero. The right-hand plot of figure~\ref{fig:lesterlallyTopdiff} compares this implementation's computed $\mttwo$ values with that of the LN algorithm, demonstrating excellent agreement to a high degree of precision.

\begin{figure}
\input{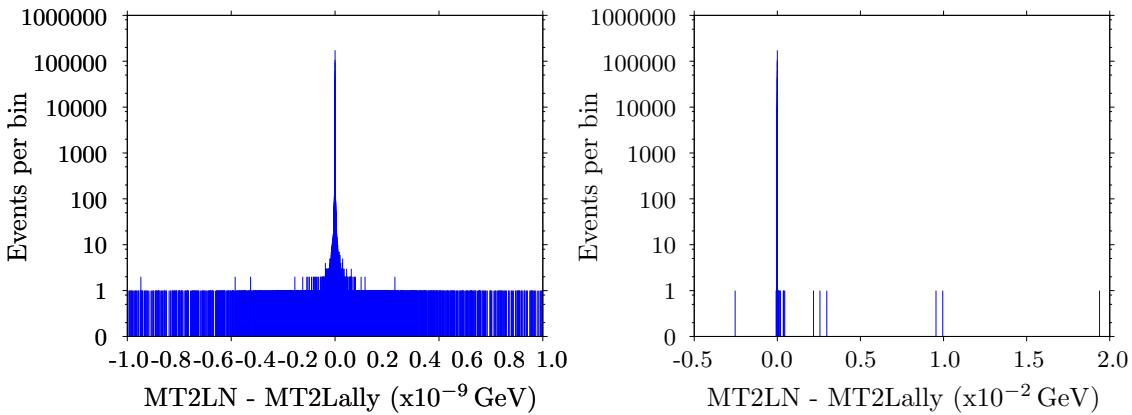}
\caption{These plots show the difference between the $\mttwo$ values calculated with the proposed implementation and that due to LN for one million (balanced) events which contain a mixture of massless, near-massless and massive but small ($< 10\GeV$) visible and invisible daughter particles (the left hand plot uses the same range as the corresponding plot in figure~\ref{fig:lesterlallyTopdiff} for ease of comparison, the right hand plot shows the full distribution). Although demonstrating again how well the two algorithms agree - better than $\sim10^{-12}\GeV$ for 93\% of the events - there is clearly a broader peak and larger tail compared with the ${t}\bar{t}$ $\mttwo$ distribution in figure~\ref{fig:lesterlallyTopdiff}. Note also the handful of events with a difference of $\sim10^{-2}\GeV$.}
\label{fig:lesterlally300slepdiff}
\end{figure}

A more challenging class of events for the calculation of $\mttwo$ would be for those on the boundary between elliptical and parabolic behaviour (i.e. massless or near-massless events). To produce a suitable dataset, the toy MC was used to produce decays from a pair of hypothetical $300 \GeV$ sparticles produced around $200 \GeV$ above threshold. Each of the visible and invisible daughter particles from the two decays had a 50\% chance of being massless, and a 50\% chance of having a mass uniformly distributed between 0 and $10 \GeV$, independent of the masses of the other daughter particles in that event. Thus a dataset of massless, near-massless and (small but) massive events was generated to investigate the proposed algorithm's accuracy in the crossover region between elliptical and parabolic behaviour\footnote{This set of events would thus be deliberately similar to those analysed in Figure 4 of LN's paper.}.

Figure~\ref{fig:lesterlally300slepdiff}  shows the new algorithm's computed $\mttwo$ values subtracted from those obtained using the LN algorithm for one million events\footnote{Note only \textit{balanced} events were used in this comparison, unbalanced events being trivial to calculate.} generated as just described. In the left hand plot the range shown has been kept the same as that for the related plot in figure~\ref{fig:lesterlallyTopdiff} to more easily compare them. For the vast majority of events ($\sim 93\%$) agreement between the two implementations is still better than $\sim10^{-12}\GeV$, however the peak is slightly broader and there is clearly more of a tail. The slightly wider peak and more extended tail is due to the greater ``volatility'' in the $\mttwo$ computation in the boundary between elliptical and parabolic behaviour; in these cusp regions a relatively smaller change to the trial parent mass, $\mu_Y$, can produce a relatively larger change in the size of the conic regions, thus limiting how precisely the value of $\mu_Y$ representing the first intersection of the conic regions can be determined i.e. even at machine precision, it is difficult to iterate any closer, and one has to accept a slightly less accurate determination of $\mttwo$. The right hand plot of figure~\ref{fig:lesterlally300slepdiff} shows the full distribution, and a few discrepancies of order $\sim10^{-2}\GeV$ are seen. These handful of events with the larger discrepancies were investigated more thoroughly (as were the outliers in the right hand plot of figure~\ref{fig:lesterlallyTopdiff}) and their $\mttwo$ values were determined by other means. These events it seems are generally those on the cusp of being unbalanced; it was found that the proposed implementation's computed values were always closer to the true value of $\mttwo$ than those due to LN's implementation in these ``quasi-unbalanced'' situations (see Appendix \ref{sec:outlieranalysis} for a detailed analysis of two of the larger outliers). It is worth remembering however that the events in figure~\ref{fig:lesterlally300slepdiff} were artificially created to be ``difficult'', and thus even with the slightly broader distribution of agreement, both implementations still represent excellent accuracy in the field of $\mttwo$ computation.

\subsection{Speed}
\label{ssec:algspeed}

The key motivation behind the proposed implementation was to find not only a more accurate way to compute $\mttwo$ but, as importantly, as fast a method as possible. As previously mentioned, given that root-finding algorithms have benefited from intense scrutiny over a very long period of time, and should therefore represent some of the most efficient numerical techniques available, the assumption has been that if the calculation of $\mttwo$ could be reduced to a root-finding problem, then this would be a sensible approach to trying the achieve the goal of improved speed. 

It is difficult to assert a definitive speed for an algorithm given the myriad of hardware and software architectures upon which it might be run; as the LN implementation and this new algorithm represent two of the most accurate methods available for computing $\mttwo$, the results presented here will instead focus on the relative performance of the two implementations when calculating $\mttwo$ to a precision of order $\sim10^{-12}\GeV$\footnote{The programs used for this speed comparison were compiled using the GCC g\texttt{++} compiler, with -03 optimisations turned on, and were run within the Cygwin linux emulator on a 2.4Ghz Intel Core i5 laptop running Windows 7.}.

\begin{figure}
\input{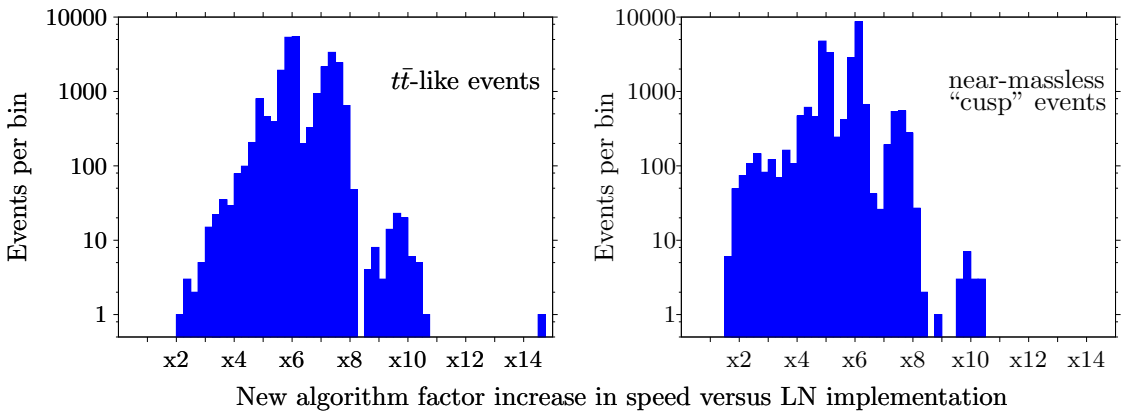}
\caption{These plots show the distribution in the relative increase in speed for the new algorithm compared with the LN implementation. The left hand plot is for the ${t}\bar{t}$ decay events; the right hand plot for the massless and near-massless events on the cusp between elliptical and parabolic behaviour.}
\label{fig:lesterlallyTopTiming}
\end{figure}

The authors of \cite{Lester:2014yga} suggest the LN algorithm when running on a modest laptop has evaluation rates of $\sim200$kHz when evaluating $\mttwo$ to a precision of $\sim10^{-12}\GeV$; this general speed, applicable to the vast majority of events evaluated using the LN implementation, was confirmed in the analysis performed herein\footnote{Note the LN algorithm was always run with the deci-section optimisation feature enabled.}.  The left hand plot of figure~\ref{fig:lesterlallyTopTiming} shows the speed improvement for the new algorithm over the LN implementation for 25,000 ${t}\bar{t}$ decay events as previously described (where $\mttwo$ is evaluated 100,000 times for each event). The results show that the new algorithm can calculate $\mttwo$ in excess of 10 times faster than the LN implementation for some types of event, that it is never less than around twice as fast, and that the average speed increase is around 6-7 times faster. There are three distinct regions in the distribution: the region around 10 times faster are for events which find the correct root first time using the Newton-Raphson method after only a handful of iterations; the region around 7 times are the ``typical'' events which find the root first time with around 10 iterations; finally the more complex region below about 6 times is mostly where the Regula Falsi method is required, and the speed of the algorithm is determined by how quickly the revised upper and lower bounds can be found prior to using the Regula Falsi method. The right hand plot of figure~\ref{fig:lesterlallyTopTiming} shows the speed increase using the more ``difficult'' massless and near-massless events on the boundary of elliptical and parabolic behaviour (as previously described); here 25,000 events were again evaluated 100,000 times each. Not surprisingly there is a slight reduction in relative speed: there are fewer events in the 10 times faster region, and more in the Regula Falsi region below 6 times, however the average speed is only marginally worse at around 5-6 times faster. Thus the conclusion with respect to computational speed for the proposed algorithm is that on a standard PC one
would expect maximum evaluation rates of $\sim2$MHz, and average evaluation rates in excess of $1$MHz.

Finally, it is worth noting that as the bisection method of the LN implementation has linear convergence, there is a corresponding increase in speed when calculating $\mttwo$ to a lower precision (the authors quote a speed increase of approximately three when the precision required goes from $\sim10^{-12}\GeV$ to $\sim10^{-3}\GeV$). However the implementation proposed herein operates with mostly quadratic (at worst superlinear) convergence (essentially meaning that when the algorithm gets near to a root, it very rapidly homes in on its true value); this means that significantly higher precision is achieved in the algorithm's last few iterations, and so in contrast to the LN implementation, there is not a significant speed advantage where a lower precision is requested.

\section{Conclusions}

A high-precision algorithm for the calculation of $\mttwo$ for both symmetric and asymmetric events has been developed which, as well as demonstrating very good stability throughout the $\mttwo$ parameter space, has been shown to be at least as accurate as currently available algorithms and can compute $\mttwo$ up to ten times faster than any previous implementation, with average evaluation rates in excess of $\sim1$MHz on a modest PC, to a precision of the order of $\sim10^{-12}\GeV$ (comparable to the most precise existing implementation due to \cite{Lester:2014yga}). It is hoped that this new implementation will be useful in situations where rapid and/or multiple computations of $\mttwo$ are required. A single C\texttt{++} header file containing the full algorithm discussed herein is available as part of the first arXiv submission of this paper.

\section{Acknowledgments}
The author is greatly indebted to Chris Lester for providing an early copy of his algorithm, a copy of his toy MC event generator, and for numerous helpful discussions. This assistance, allowing the proposed implementation to be thoroughly validated, has been invaluable.

\appendix 
\section{Extending the validity of the method to parabolae}
\label{sec:validityconics}

As discussed in the main body of the paper, the implementation proposed herein relies on the work summarised in \cite{choi2006continuous}\footnote{As does LN's implementation, since both derive ultimately from the idea to use properties of the characteristic polynomial first proposed by \cite{wang2001algebraic}.}, however the proofs presented in that paper focus specifically on ellipses (for reasons explained previously). It would be useful to investigate if there is any reason to believe that the five key lemmas leading to Theorem 6 of \cite{choi2006continuous} cannot be extended to cover parabolae\footnote{The following discussion is not attempting to be a formal proof that Theorem 6 can be extended to parabolae. However, when read in conjunction with a knowledge of the proofs contained in \cite{choi2006continuous}, it is hoped the logical arguments put forward in this appendix are sufficient to give the reader comfort that it would be relatively straightforward to formally extended Theorem 6 to cover the intersection of parabolae.}.

The five lemmas are as follows (note a distinction is made between the boundary of an elliptical region (i.e. $X^T A X = 0$) and the interior of an elliptical region (i.e. $X^T A X < 0$), and only the part of lemma 5 relevant to the computation of $\mttwo$ is discussed):

\begin{description}
\item[Lemma 1] The characteristic polynomial derived from two elliptical regions, $f(\lambda) = 0$, either has three positive roots, one positive and two negative roots, or one positive and a pair of complex conjugate roots;
\item[Lemma 2] If the interiors of two elliptical regions do not intersect, then $f(\lambda) = 0$ has a negative root;
\item[Lemma 3] If the interiors of two elliptical regions intersect, then any real root of $f(\lambda) = 0$ is positive;
\item[Lemma 4] If two elliptical regions touch externally, then $f(\lambda) = 0$ has a negative double root;
\item[Lemma 5] If $f(\lambda) = 0$ has a negative double root, then the elliptical regions $\mathcal{P} : X^T A X \leq 0$ and $\mathcal{Q} : X^T B X \leq 0$ touch each other externally.
\end{description}

Lemma 1 and the relevant part of lemma 5 (the second paragraph for those wishing to review it) do not distinguish between the type of conic and so the proofs are applicable to any conic (indeed the relevant part of lemma 5 is actually proven for general conics). Lemma 4 also does not distinguish between the type of conic (however it does depend on lemma 2). Lemma 3 does not technically distinguish between the type of conic, but it does require there to be a point on the plane exterior to both conics, which for two finite {\it bounded ellipses} is self-evident. This is not so clear-cut for parabolae, as two parabolic regions with their vertices infinitely displaced from each other (e.g. one vertex at $-\infty$ the other at $+\infty$) could in principle together cover the plane. However for any parabolic region describing the allowed transverse momenta in the computation of $\mttwo$, at least one of these parabolae will, for physical reasons, have a finite vertex. This therefore would ensure that there is always a point on the plane exterior to both and thus allows lemma 3 to be trivially extended to parabolae. In order to extend Theorem 6 to parabolic regions we are therefore left with only needing to extend lemma 2, which is slightly more involved and is specifically proven for elliptical disks only. We shall take a brief look at the key points involved in the proof and then see how they might be extended to parabolae.

\begin{figure}
\centering
\includegraphics[width=10cm]{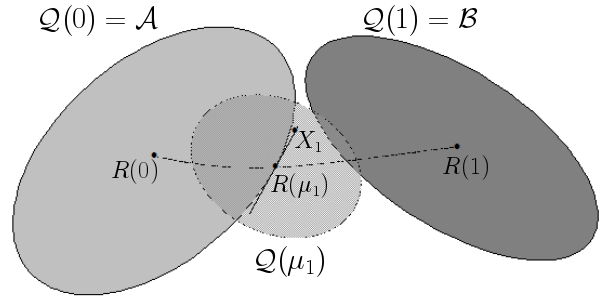}
\caption{Configuration for Lemma 2 (reproduced from \cite{choi2006continuous}).}
\label{fig:Lemma2Ellipse}
\end{figure}

The situation is as depicted in figure~\ref{fig:Lemma2Ellipse}, where we have two elliptical regions that do not intersect (note however that they are allowed to touch on their external boundary and this would not affect the proof). The substitution $\lambda = (\mu - 1)/\mu$ (which maps $\mu \in [0,1]$ to $\lambda \in [-\infty, 0]$) is made which transforms the characteristic equation $f(\lambda) = det(\lambda A - B) = 0$ to $g(\mu) \equiv det((1 - \mu) A + \mu B) = 0$. A new conic can be made, $Q(\mu) \equiv (1 - \mu) A + \mu B$, and it is straightforward to see that $Q(0)$ is equivalent to the elliptical region $A$ and $Q(1)$ is equivalent to elliptical region $B$; the proof then shows that $\mathcal{Q}(\mu)$ must also represent an elliptical region as $det(Q_{2,2}) > 0$ (given that $\mathcal{A}$ and $\mathcal{B}$ were defined to be elliptical regions so that  $det(A_{2,2}) > 0$ and  $det(B_{2,2}) > 0$). The crux of the proof is then to show that there must exist a point $X_1$ that is interior to an elliptical region $\mathcal{Q}(\mu_1)$ (where $0 < \mu_1 < 1$) but which is exterior to both $\mathcal{A}$ and $\mathcal{B}$.

\begin{figure}
\centering
\includegraphics[width=10cm]{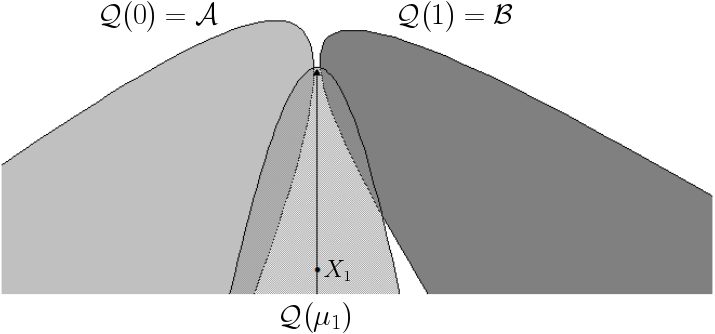}
\caption{Configuration for Lemma 2 extended to parabolae.}
\label{fig:Lemma2Parabola}
\end{figure}

To extend this to parabolae we can proceed in a similar way. However now instead of defining  $\mathcal{A}$ and $\mathcal{B}$ to be elliptical regions, we allow them to be either elliptical or parabolic regions i.e. by allowing $det(A_{2,2}) \geq 0$ and  $det(B_{2,2}) \geq 0$. This then leads, by the same argument, to $\mathcal{Q}(\mu)$ representing {\it either} an elliptical or parabolic region i.e. $det(Q_{2,2}) \geq 0$. If the conic regions are elliptical we already have a proof, if they are parabolic then we have the situation depicted in figure~\ref{fig:Lemma2Parabola}. Here it is arguably easier to show that  there must exist a point $X_1$ that is interior to the parabolic region $\mathcal{Q}(\mu_1)$ but which is exterior to both parabolae $\mathcal{A}$ and $\mathcal{B}$. For example we can translate and rotate $\mathcal{A}$ and $\mathcal{B}$ (note such affine transformations will not alter the roots of the characteristic equation) so that they are separated by the y-axis (or if the two conics touch on their external boundary, the y-axis can be made tangent to both conics at the intersection point). We then have $\mathcal{A}$ centered at $(-\infty, -\infty)$,  $\mathcal{B}$ centered at $(\infty, -\infty)$, and define $\mathcal{Q}(\mu_1)$ to be the parabola centered at $(0, -\infty)$. It is clear that there is a point $X_1$ located along the y-axis somewhere between the vertex and center of $\mathcal{Q}(\mu_1)$ (but excluding the tangent point with both parabolae if they are touching externally) which is internal to $\mathcal{Q}(\mu_1)$ but external to $\mathcal{A}$ and $\mathcal{B}$. This simple argument is enough to allow the proof of lemma 2 to be concluded as per the elliptical case and thus extend both lemma 2 and thus Theorem 6 to parabolae.

Although the above is not a {\it formal} proof that these lemmas can be extended to the parabolic case (and thus that Theorem 6 is valid for parabolae), it should give significant comfort that there do not seem to be any obvious impediments to doing so and that a formal extension would be a reasonably trivial affair. In their paper, LN also recognise the lack of a formal proof (though one of the authors (CGL) similarly conjectures that it should be extendable to parabolae), and the above should hopefully give users of the LN implementation greater comfort that this conjecture is likely to be true.

\section{Analysis of discrepancies in the evaluation of $\mttwo$ between the Lester-Nachman and the proposed implementations}
\label{sec:outlieranalysis}

In section~\ref{ssec:algaccuracy} we saw in figure~\ref{fig:lesterlally300slepdiff} that the LN implementation and the one proposed here did not always agree on the $\mttwo$ value, and there were a handful of discrepancies as large as $\sim10^{-2}\GeV$ (and some smaller discrepancies of order $\sim10^{-9}\GeV$ in figure~\ref{fig:lesterlallyTopdiff}). A large number of the events which give rise to discrepancies between the two algorithms have been looked at in detail and, as they fall into two main ``quasi-unbalanced'' categories, in this appendix we will look in more detail at just two events (chosen to represent the two categories); these events are two of the larger differences in $\mttwo$ value visible in figure~\ref{fig:lesterlally300slepdiff}, with discrepancies of $0.00954 \GeV$ and $-0.00254 \GeV$ respectively. 

The first event has the following parameters (notation is as before with subscripts 1 and 2 indicating sides of an event): $m_1 = 0.00473856926 \GeV$, $p_{1x} = 110.43799970914 \GeV$, $p_{1y} = -213.46687262192 \GeV$, $m_2 = 0.0 \GeV$, $p_{2x} = -20.28455035002 \GeV$, $p_{2y} = \\235.76522546534 \GeV$, $\slashed{p}_{T}^{x} = 111.30684472357 \GeV$, $\slashed{p}_{T}^{y} = 37.47049084405 \GeV$, $\mu_{N1} = \mu_{N2} = 0.0 \GeV$; the correct $\mttwo$ value for this event is $0.007098115 \GeV$. The parameters for the second event are:   $m_1  =  0.21164596081 \GeV$,  $p_{1x}  =  52.78735611764 \GeV$,  $p_{1y}  = -34.61581597866 \GeV$, $m_2 = 0.0 \GeV$, $p_{2x} = -27.12567045837 \GeV$, $p_{2y} = \\-8.57910177494 \GeV$, $\slashed{p}_{T}^{x} = -245.80036782403 \GeV$, $\slashed{p}_{T}^{y} = -77.06353868081 \GeV$, $\mu_{N1} = \mu_{N2} = 0.0 \GeV$; the correct $\mttwo$ value for this event is $0.214304220 \GeV$.

\begin{figure}
\centering
\includegraphics[width=15.0cm]{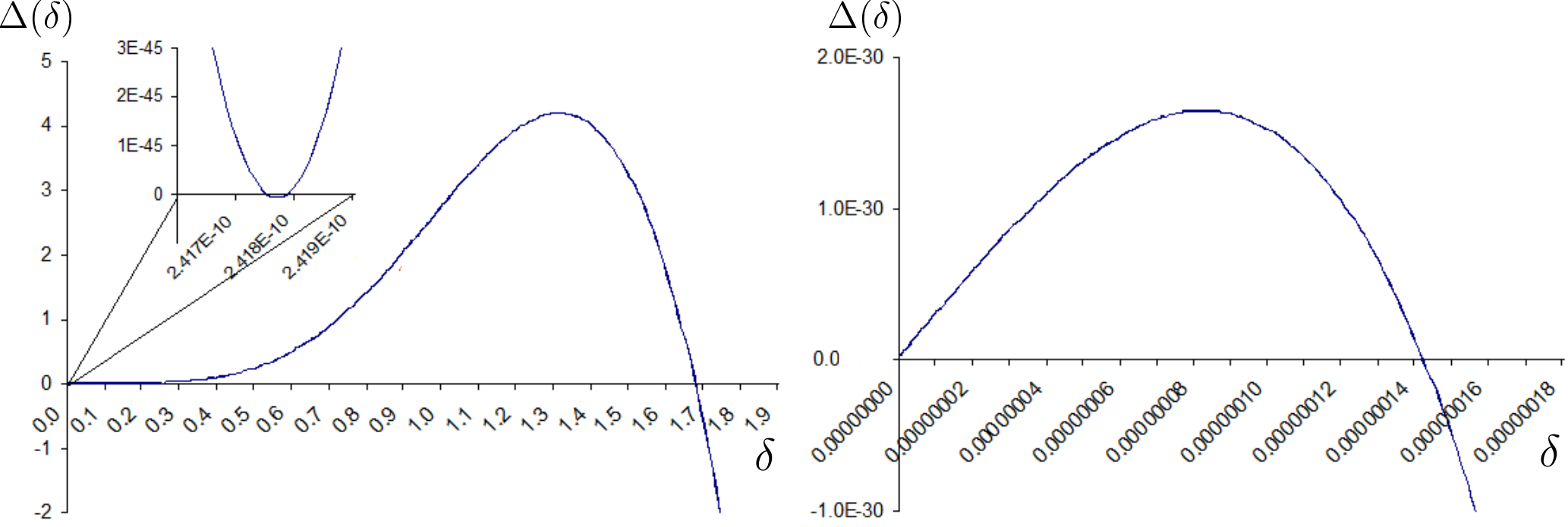}
\caption{Discriminant functions for the two events, both showing a zoomed-in section near the origin where the lowest root can be found. For the first event the lowest root ($\delta  = 2.4174985\times10^{-10}\GeV$) equates to a ``trial'' parent mass, $\mu_Y$, value of $0.007098115 \GeV$; for the second event the lowest root ($\delta  = 1.4207550\times10^{-7}\GeV$) equates to a $\mu_Y$ value of $0.214304220 \GeV$.}
\label{fig:DeltaFuncsOutliers}
\centering
\includegraphics[width=15.0cm]{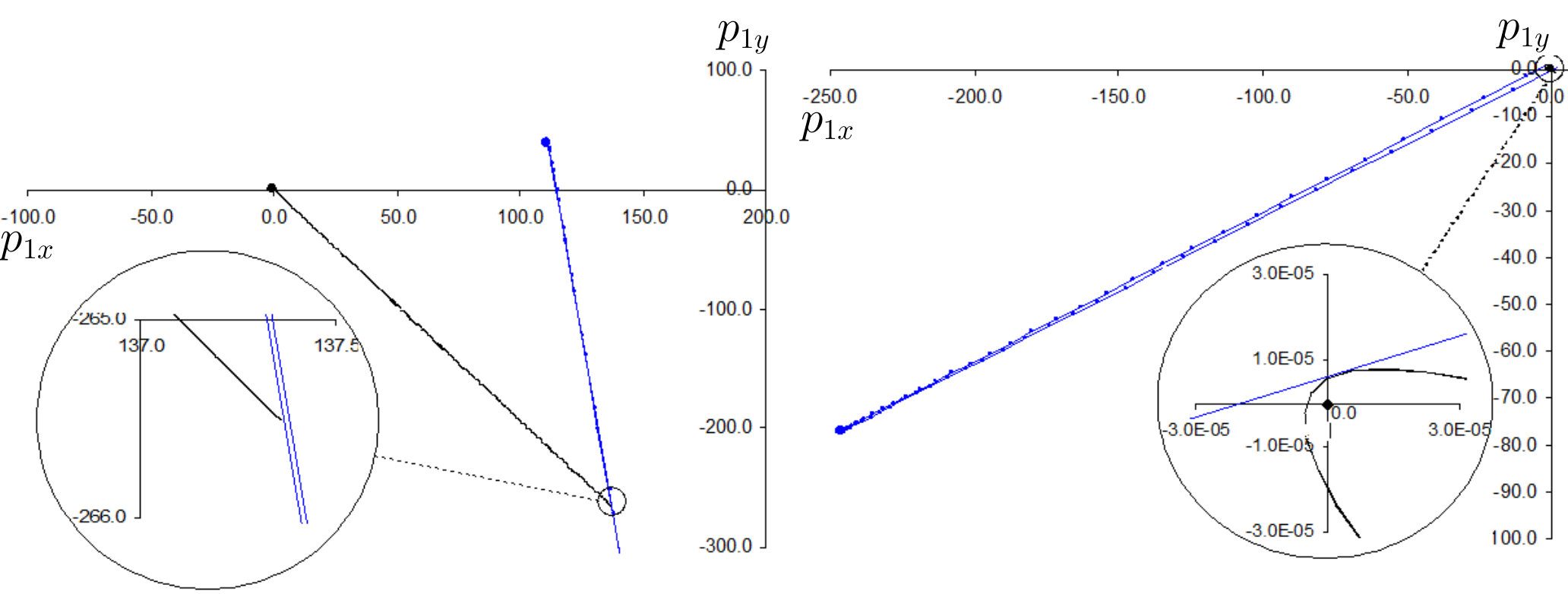}
\caption{Conic regions for the two events. The zoomed-in section shows that the first intersection of side 1's ellipse with side 2's parabola occurs at the trial parent masses obtained from the lowest root of the discriminant function, confirming the $\mu_Y$ values as the correct $\mttwo$ values for the events.}
\label{fig:LallyConicIntersections}
\end{figure}

Figure~\ref{fig:DeltaFuncsOutliers} shows the cubic characteristic polynomial discriminant functions for these two events; the first has five real positive roots (only three of which are visible in the plot) and the second event has only one real root; the two lowest roots for event 1 (visible in the zoomed-in section) are very close together and small in value (i.e. close to zero\footnote{Indeed if the small mass of side 1 of this ``quasi-unbalanced'' event was actually zero, it would be unbalanced and the $\mttwo$ value would be exactly zero.}).

 Figure~\ref{fig:LallyConicIntersections} shows the two conical regions for each event, both evaluated at their respective trial parent mass values as obtained from the lowest root of the discriminant functions shown in figure~\ref{fig:DeltaFuncsOutliers}. Side 1 of both events is a (very thin) ellipse which first manifests at the origin, at the respective event's side 1 mass value. Side 2 for both events is a (very thin) parabola whose vertex is the event's respective missing transverse momenta  $\slashed{p}_{T}^{x}$ and  $\slashed{p}_{T}^{y}$ values\footnote{Note that event 2 is also ``quasi-unbalanced'' in the sense that the origin (where side 1's ellipse first manifests) is only just outside of the parabola; were the origin on or inside the parabola when the ellipse first manifested, the event would be unbalanced and the $\mttwo$ value would be simply the mass of side 1, $m_1$.}.  It is clear from the zoomed-in sections of the figure that the two conic regions do indeed first intersect at the respective trial mass values, thus confirming that these $\mu_Y$ values are the correct $\mttwo$ values for the events.
\begin{figure}
\centering
\includegraphics[width=15.0cm]{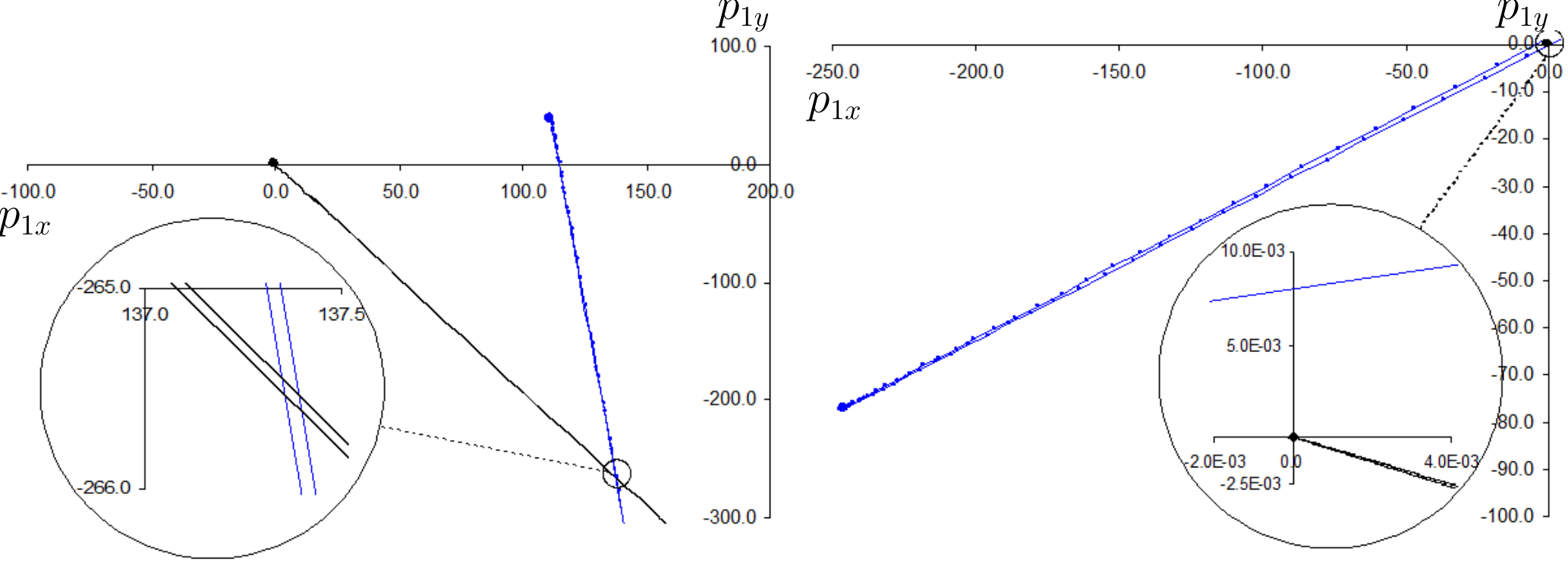}
\caption{Conic regions for the two events using the $\mttwo$ values as calculated by the LN algorithm. It is clear that event 1's value is too high as side 1's ellipse has grown well beyond the point of initial intersection with the parabola of side 2; for event 2 the value is too small as neither the ellipse has grown, nor the parabola widened, sufficiently for first intersection.}
\label{fig:LNConicIntersections}
\end{figure}

For completeness, figure~\ref{fig:LNConicIntersections} shows the conic regions at the $\mttwo$ values evaluated by the LN algorithm ($0.016643073 \GeV$ and $0.211763085 \GeV$ respectively); it is clear that the $\mttwo$ value for event 1 is too large as the ellipse has grown well beyond the point of initial intersections with the parabola; for event 2, the value is too low as the parabola needs to widen further, and the ellipse expand further, before the conical regions will first intersect.

\bibliographystyle{JHEP-withSlacCitation}
\bibliography{paper}

\end{document}